\documentclass{sf2a-conf2011}
\usepackage{graphicx}
\usepackage{hyperref}
\usepackage[]{natbib}
\usepackage[cyr]{aeguill}
\usepackage{epstopdf}
\usepackage{amsmath}
\usepackage{booktabs}

\def\BibTeX{{\rm B\kern-.05em{\sc i\kern-.025em b}\kern-.08em
    T\kern-.1667em\lower.7ex\hbox{E}\kern-.125emX}}
\bibpunct{(}{)}{;}{a}{}{,}  
\newcommand{\lp}{ \left(}
\newcommand{\rp}{ \right)}
\newcommand{\Et}{{\cal E}_{t}}
\newcommand{\Fst}{{\cal F}_{\text{St}}}

\newcommand{\Fku}{{\cal F}_{\text{Ku}}}
\newcommand{\alpst}{\alpha_{\text{St}}}
\newcommand{\alpku}{\alpha_{\text{Ku}}}
\newcommand{\gradad}{\nabla_{\text{ad}}}
\DeclareMathOperator{\sign}{sign}

\begin{document}

\TitreGlobal{SF2A 2011}


\title{Nonlinear simulations of the convection-pulsation coupling}

\runningtitle{Nonlinear simulations of the convection-pulsation coupling}

\author{T. Gastine}\address{Max-Planck-Institut f\"ur Sonnensystemforschung, 
Max-Planck-Strasse 2, 37191 Katlenburg-Lindau, Germany}

\author{B. Dintrans}\address{IRAP, CNRS-Universit\'e de Toulouse, 
14 avenue Edouard Belin, F-31400 Toulouse, France}

\setcounter{page}{237}

\index{Gastine, T.}
\index{Dintrans, B.}


\maketitle


\begin{abstract}
In cold Cepheids close to the red edge of the classical instability strip, a 
strong coupling between the stellar pulsations and the surface convective 
motions occurs. This coupling is by now poorly described by 1-D models of 
convection, the so-called ``time-dependent convection models" (TDC). The 
intrinsic weakness of such models comes from the large number of unconstrained 
free parameters entering in the description of turbulent convection. A way to 
overcome these limits is to compute two-dimensional direct simulations (DNS), 
in which all the nonlinearities are correctly solved. Two-dimensional DNS of the
convection-pulsation coupling are presented here. In an appropriate 
parameter regime, convective motions can actually quench the radial pulsations 
of the star, as suspected in Cepheids close to the red edge of the instability 
strip. These nonlinear simulations can also be used to determine the limits and 
the relevance of the TDC models.
\end{abstract}



\section{Introduction}
The cold Cepheids located close to the red edge of the classical instability
strip have a large surface convective zone that affects their
pulsation properties \citep[e.g. the reviews
of][]{Gautschy-Saio,buchler-cepheid-review}. The first calculations, that
assumed \textit{frozen-in} convection, predicted a cooler red edge than the
observed one. Indeed, as already stated by \cite{Baker-Kippenhahn}, a
non-adiabatic treatment of the convection-pulsation coupling is mandatory to
predict the red edge location with a better accuracy.

Several time-dependent convection (TDC) models were therefore developed to
address this coupling \citep[e.g.][]{St, K, Xiong89} and succeeded in
reproducing the correct location of the red edge, despite their disagreements
with the physical origin of the mode stabilisation \citep[e.g.][]{Bono,
YKB98,Dupret1}.

However, all these formulations involve many free and degenerate parameters
\citep[e.g. the seven dimensionless $\alpha$ coefficients used by][]{YKB98}
that are either fitted to the observations or hardly constrained by theoretical
values. Nevertheless, another way to tackle this problem is to compute 
2-D and 3-D direct numerical simulations (DNS) that correctly take into account
the nonlinearities involved in this coupling. Results of such pioneering 2-D
nonlinear simulations of the convection-pulsation coupling are presented in the
following.

\section{The convection-pulsation coupling}
Our system corresponds to a local zoom around an ionisation region responsible
for the driving of the acoustic modes excited by the $\kappa$-mechanism. It is
composed of a 2-D cartesian layer filled with a monatomic and perfect gas. The
opacity bump associated with this ionisation zone is modelled by a
temperature-dependent radiative conductivity profile $K(T)$ \citep[for further
details, see][]{paperI}. In addition to the $\kappa$-mechanism, the
conductivity profile is deep enough to locally get a superadiabatic temperature
gradient, meaning that convective motions develop there, according to
Schwarzschild's criterion. A strong coupling between convection and the
acoustic oscillations therefore develops.

The hydrodynamical equations are then advanced in time
with the high-order, finite-difference \textit{Pencil Code}\footnote{See 
\url{http://www.nordita.org/software/pencil-code/} and
\cite{Pencil-Code}.}, which is fully explicit except for the radiative
diffusion term that is solved implicitly thanks to a parallel
alternate direction implicit (ADI) solver of our own \citep{paperI}. The
simulation box spans about $10\%$ of the star radius around the
ionisation region. In order to ensure that both the thermal relaxation and 
nonlinear
saturation of the $\kappa$-mechanism are achieved, the simulations are computed
over more than 4000 days \citep[corresponding roughly to 1500 periods of
oscillation, see][]{paperIII}.

\begin{figure}
 \centering
 \includegraphics[width=0.8\textwidth,clip]{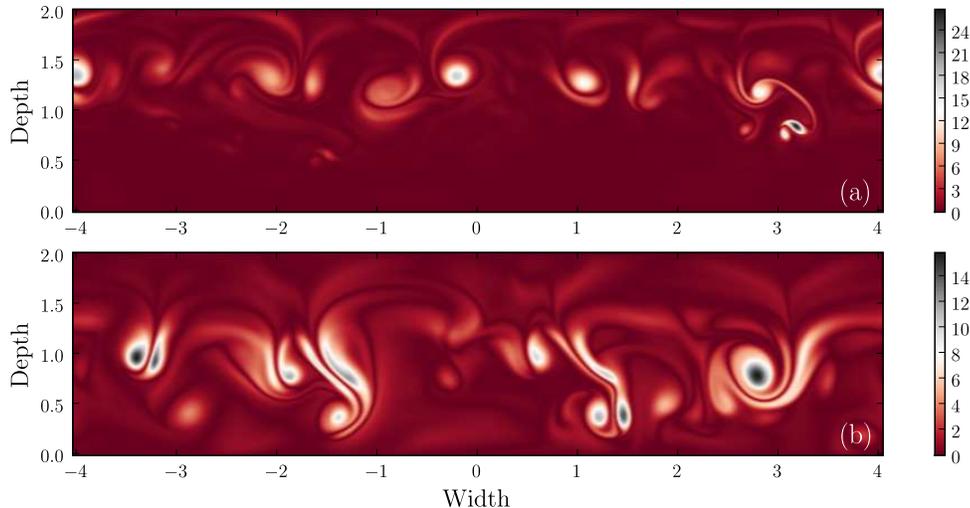}
 \caption{Snapshot of the modulus of the vorticity field $|\vec{\nabla}\times
\vec{u}|$ in the G8 (\textbf{a}) and in the G8H8 simulations
(\textbf{b}).}
 \label{snapshot}
\end{figure}

Figure~\ref{snapshot} displays a snapshot of the vorticity field
for two simulations discussed in \cite{paperIII}, namely G8 (upper panel)
and G8H8 (lower panel).
This vorticity field highlights the convective motions that are approximately
localised in the middle of the layer, where the radiative conductivity is
minimum. Differences in the typical length-scale of convection are
noticeable between these two DNS: convective eddies are smaller scale in the
G8 simulation than in the G8H8 one. Accordingly, the overshooting of convective
elements into the lower stably stratified layer is also more pronounced in the
latter simulation. 

\begin{figure}
 \centering
 \includegraphics[width=0.45\textwidth,clip]{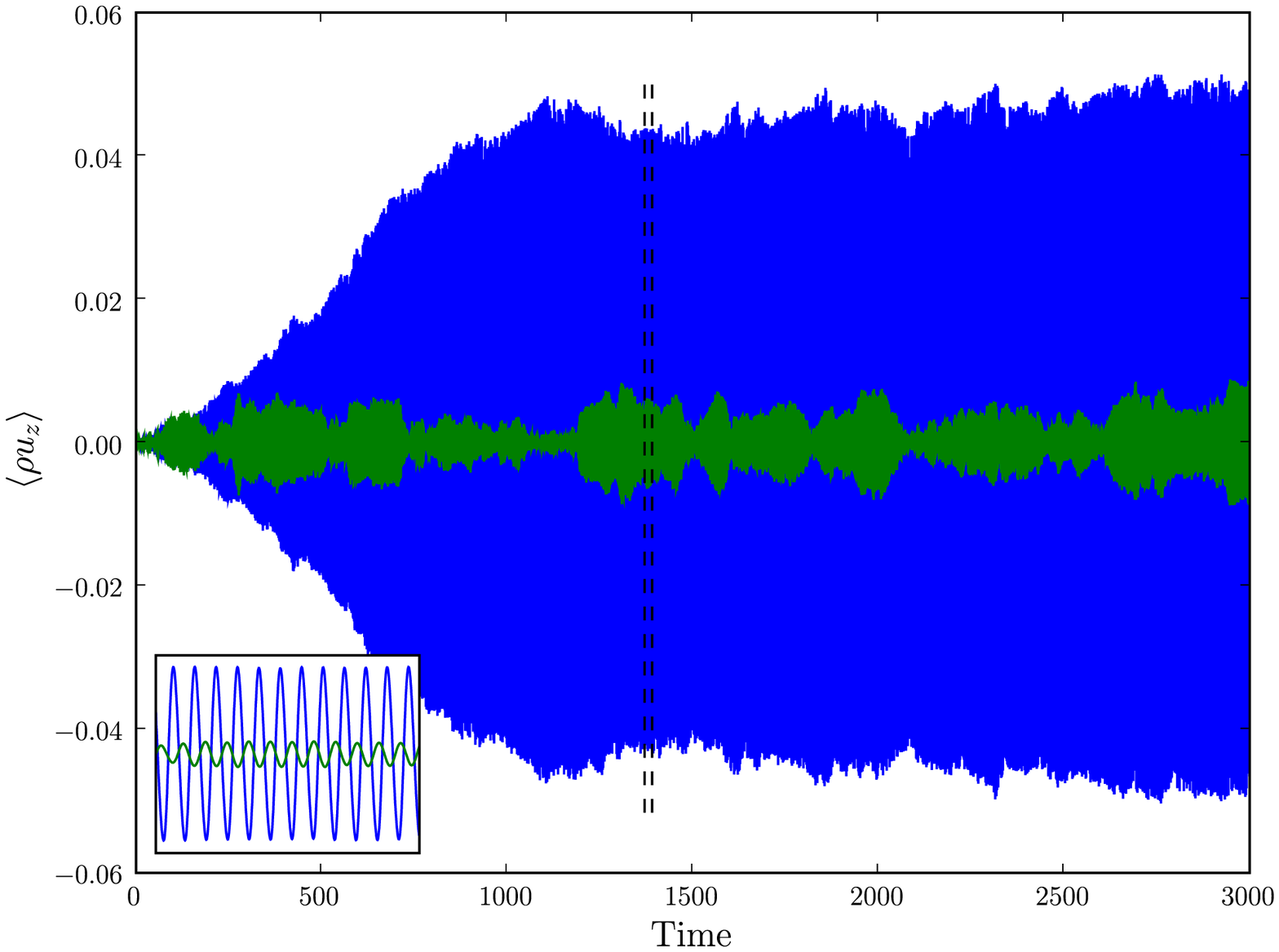}
 \includegraphics[width=0.45\textwidth,clip]{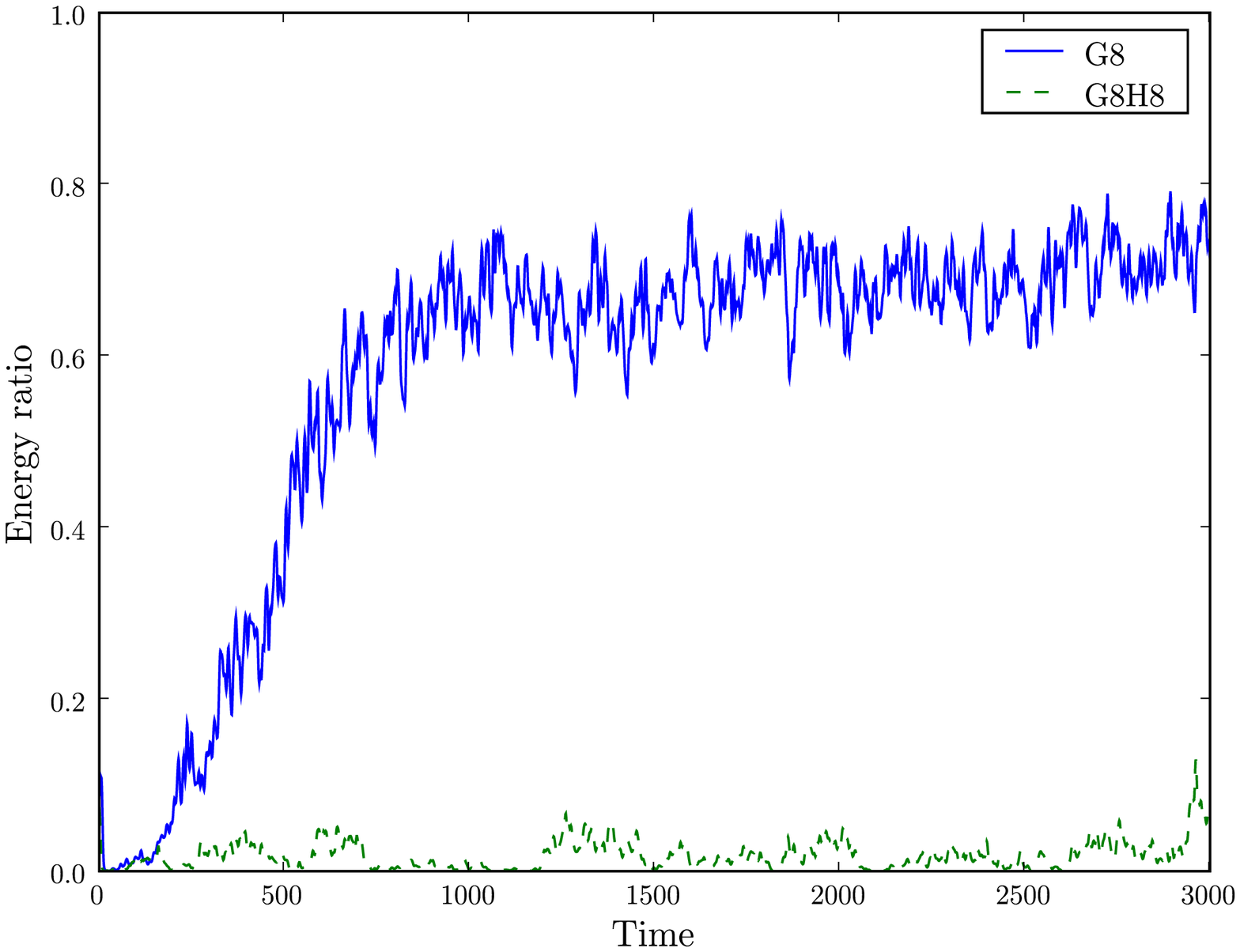}
 \caption{\textit{Left panel:} Temporal evolution of the mean vertical mass
flux $\rho u_z$ for the two simulations G8 (solid blue line)
and G8H8 (solid green line). The two vertical dashed black lines define the
boundaries of the zoom displayed in the bottom left corner. \textit{Right
panel:} temporal evolution of the energy contained in acoustic modes
normalised by the total kinetic energy for the
two simulations G8 (solid blue line) and G8H8 (dashed green line).}
 \label{ts}
\end{figure}

Beyond these qualitative differences, a good way to compare these simulations is
to study the temporal evolution of average quantities, such as the vertical mass
flux $\rho u_z$. Indeed, as we are considering simulations with both convective
motions and oscillations of acoustic modes, it is relevant to use
a simple diagnostic that roughly separates their relative contributions. 
Because the convective plumes have both ascending and descending motions, the
average vertical mass flux filters out their contribution and is
therefore a good proxy of the amplitude of the acoustic modes. The left
panel of Fig.~\ref{ts} therefore displays the temporal evolution of $\rho u_z$
for the two simulations discussed before. An oscillatory behaviour is observed
in both cases due to the radial oscillations of the
fundamental acoustic mode excited by $\kappa$-mechanism. In the
G8 simulation, the amplitude first grows exponentially until
reaching the nonlinear saturation regime. At first glance,
this time evolution looks very similar to what has been already observed in
our former purely radiative simulations done in \cite{paperII}, that is, a 
linear growth of the
amplitude and a saturation at a well-defined value. In contrast, the dynamics of
the G8H8 simulation differ radically from the previous one
as the amplitude remains weak and is highly modulated over time. No clear
nonlinear saturation is observed in this case, meaning that the
acoustic oscillations are more influenced by convective motions than in the
previous DNS.

To separate more precisely the relative contributions of the acoustic modes and
the convective motions to the energy budget, the velocity field of each
simulations is projected onto an acoustic subspace built from normal eigenmodes
\citep[see][]{Bogdan1993}. Thanks to this formalism, it is possible to extract
the time evolution of the kinetic energy contained in each acoustic
mode found in the nonlinear simulation. The right panel of Fig.~\ref{ts} 
displays the time evolution of the energy contained in the fundamental
acoustic mode normalised by the total kinetic energy. In the G8 simulation, the
acoustic energy linearly increases until its nonlinear saturation, in
a similar way to what has been previously observed with the time evolution of
$\rho u_z$. Once this saturation is reached, $70\%$ of the kinetic energy is
contained in the radial oscillations of the fundamental acoustic mode, while
the remaining is in the convective plumes. In other words, the acoustic
oscillations are not affected much by the convective motions in this simulation.
In contrast, this acoustic energy ratio remains very
weak in the other simulation. Despite some transient increases
during which non-trifling values ($\simeq 10\%$) are obtained, the
average ratio is less than $5\%$, and convective motions contains the
bulk of the kinetic energy. In this case, the radial oscillations
excited by the $\kappa$-mechanism are thus quenched by convective plumes.
This situation is relevant to the physics of Cepheids close to the red edge of
the instability strip, where the unstable acoustic modes are supposed to be
damped by the surface convective motions. This convective quenching of the
acoustic oscillations may be the direct signature of the different density
contrasts in the G8 and the G8H8 simulations: in fact, weaker stratification (as
in G8H8) leads to bigger vortices (see Fig.~\ref{snapshot}b), meaning that the
energy is contained in larger convective structures. In our DNS, the
amplitude of
the $\kappa$-mechanism seems to be controlled by the screening effect due to
these large convective vortices \citep{paperIII}.

\section{Limits of time-dependent convection models}

The nonlinear simulations of the convection-pulsation coupling, where the
acoustic modes strongly modulate the convective motions over time (as in the
G8 simulation) are also good candidates to test and compare the relevance of
different prescriptions of 1-D time-dependent convection (TDC) models. We focus
here on two popular formulations widely used in Cepheids models, namely the
TDC models of \cite{St} and \cite{K}. In these formulations, a single
equation for the turbulent kinetic energy $\Et$ is added to the classical
mean-field equations and the main second-order correlations, such as the
convective flux, are expressed as a function of $\Et$ only:
  
\begin{equation}
\left\lbrace
\begin{aligned}
 \Fst(z,t) & = \alpst \dfrac{A}{B} \Et \sign(\nabla
-\gradad)\sqrt{|\nabla -\gradad |}, \\ 
 \Fku(z,t) & = \alpku A \sqrt{\Et}\lp \nabla -\gradad \rp,
\end{aligned}
 \right.
 \label{eq:MLT}
\end{equation}
where $\nabla=d\ln T/d\ln p$, $\gradad=1-c_v/c_p$ and

\begin{equation}
\Et(z,t) =\left\langle\dfrac{{u'_z}^2}{2}\right \rangle,\
A = c_p\left\langle \rho \right\rangle \left\langle T \right\rangle
\text{ and } B= \sqrt{c_p\left\langle T \right\rangle \gradad},
\label{eq:eturb}
\end{equation}
where $p$ is the pressure, $\rho$ the density, and the brackets denote a
horizontal average. These two TDC expressions involves one
dimensionless parameter $alpha$ ($\alpst$ and $\alpku$, respectively) that is 
poorly
constrained by theory \citep{YKB98}. The nonlinear results of the G8
simulation allow us to constraint it by using the $\chi^2$-statistics so
that to obtain its optimum value for each formulation.

\begin{figure}
 \centering
 \includegraphics[width=0.45\textwidth,clip]{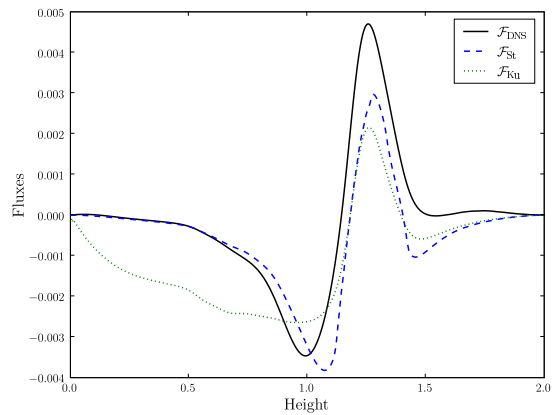}
 \caption{Mean convective flux in the G8 simulation (solid black line), compared
 with the best TDC predictions based on the models of Stellingwerf (dashed blue
line) and Kuhfu\ss~(dotted green line).}
 \label{flux}
\end{figure}

Figure~\ref{flux} compares the convective flux in the DNS with the two TDC
prescriptions computed with the best-fit values obtained for $\alpst$ 
and $\alpku$ \citep{letter}. Stellingwerf's formulation seems to
give a better agreement with the nonlinear simulation than Kuhfu\ss's one.
Indeed, the Kuhfu\ss~model overestimates the overshooting as the (negative)
convective flux remains non-negligible until the bottom of the radiative zone.
In contrast, the Stellingwerf profile accounts for the local 
penetration of convective plumes better and shows the expected exponential-like
decay in the negative convective flux when sinking in the
radiative zone. However, the two models are fairly similar in the bulk of the
convective zone, where convection is fully developed. One also notes that they
both predict a negative flux at the top of the convective zone, which is
an upper overshooting of convective motions near the surface that is
not observed in the DNS.

\section{Conclusion}
The main weakness of all theories of 1-D time-dependent convection lies in
the large number of free parameters involved in the description of convection.
New constraints must therefore be found to reduce the intrinsic degeneracy of
these models and to check the relevance of the different assumptions underlying
these parametrisations.

Nonlinear 2-D direct numerical simulations are a useful way to address the
convection-pulsation coupling that occurs in cold Cepheids. These simulations
show a variable influence of convection onto the acoustic modes excited by
$\kappa$-mechanism: (\textit{i}) either the amplitude of the acoustic modes
remains very weak and convective motions quench the oscillations ; (\textit{ii})
or the kinetic energy is mainly contained in the acoustic modes and convective
plumes are strongly modulated over time by the radial oscillations.
While the former situation is relevant to the stabilisation of the oscillations
of Cepheids close to their red edge, the latter is a good candidate to draw
the limits of current TDC recipes. Focusing on two such widely used models,
Stellingwerf's formulation is found to give a better agreement with the
nonlinear results than does Kuhfu\ss's. 

This first comparison of TDC theories with fully nonlinear results
coming from 2-D direct simulations of the convection-pulsation coupling
emphasises how DNS can be helpful to validate and improve 1-D models of
time-dependent convection.

\begin{acknowledgements}
This work was granted access to the HPC resources of CALMIP under the
allocation 2010-P1021 (\url{http://www.calmip.cict.fr}).
\end{acknowledgements}

\bibliographystyle{aa}  

%
\end{document}